\documentclass[12pt,preprint]{aastex}
\usepackage{graphicx}
\usepackage{amsmath}

\newcommand{\kms}{\ensuremath{\mathrm{km\,s^{-1}}}}

\shorttitle{Mild steepening IMF in very massive ETGs}
\shortauthors{Spiniello et al.}

\begin{document}

\title{Evidence for a mild steepening and Bottom-heavy IMF in Massive\\ Galaxies from Sodium and Titanium-Oxide Indicators}

\author{C. Spiniello\altaffilmark{1},  S.~C. Trager\altaffilmark{1},
  L.~V.~E. Koopmans\altaffilmark{1}, Y.~P. Chen\altaffilmark{1}}
\altaffiltext{1}{Kapteyn Institute, University of Groningen, PO Box
  800, 9700 AV Groningen, the Netherlands}

\begin{abstract}
We measure equivalent widths 
(EW) -- focussing on two unique features (NaI and TiO2) of low-mass stars ($\la 0.3$M$_{\odot}$) -- 
for luminous red galaxy spectra from the the Sloan Digital Sky Survey (SDSS) 
and X-Shooter Lens Survey (XLENS) in order to study the low-mass end of the initial mass function (IMF).  
We compare these EWs to those derived from
simple stellar population models computed with different IMFs, ages,
[$\alpha$/Fe], and elemental abundances.
We find that models are able to simultaneously reproduce the
observed NaD $\lambda$5895 and \ion{Na}{1} $\lambda$8190 features for lower-mass ($\sim \sigma_{*}$) early-type galaxies (ETGs) but deviate increasingly for more massive
ETGs, due do strongly mismatching NaD EWs. The TiO2 $\lambda$6230 and 
the \ion{Na}{1} $\lambda$8190 features together appear to be a powerful IMF diagnostic, with age and metallicity effects orthogonal to the effect of IMF. 
We find that both features correlate strongly with galaxy velocity dispersion. 
The XLENS ETG (SDSSJ0912+0029) and an SDSS ETG (SDSSJ0041-0914) 
appear to require both an extreme dwarf-rich IMF and a high sodium enhancement
($\mathrm{[Na/Fe]}=+0.4$).
In addition, lensing constraints on the total mass of the XLENS system
within its Einstein radius limit a bottom-heavy IMF with a power-law
slope to  $x\le 3.0$ at the $90\%$ C.L. We conclude that NaI and TiO features, in comparison with 
state-of-the-art SSP models, suggest a mildly steepening IMF from Salpeter ($dn/dm\,\propto\,m^{-x}$ with $x=2.35$) to 
$x \approx 3.0$ for ETGs in the range $\sigma = 200- 335$\,\kms . 
\end{abstract}


\section{Introduction}
When constraining the star formation, metallicity and
gas/dust content of galaxies, the initial mass function (IMF) is often assumed to be universal 
and equal to that
of the solar neighborhood (Kroupa 2001; Chabrier 2003; Bastian, Covey
\& Meyer 2010).  However, 
evidence has recently emerged that the IMF might evolve (Dav\'{e} 2008;
van Dokkum 2008) or depend on the stellar mass of the system
(e.g. Worthey 1992; Trager et al.\ 2000b; Graves et al.\ 2009; Treu et al.\ 2010; Auger et al.\ 2010b; Napolitano
2010; van Dokkum \& Conroy 2010). van Dokkum \& Conroy
(2010; hereafter vDC10) suggested that low-mass stars ($\leq
0.3\,M_{\odot}$) could be more prevalent in massive
early-type galaxies. The
increase in the mass-to-light ratio (M/L) of galaxies with galaxy mass
may thus be partly due to a changing IMF rather than an
increasing dark matter fraction, consistent with previous suggestions
(Treu et al.\ 2010, Auger et al.\ 2011, Barnab\`{e} et al.\ 2011, Dutton et al.\ 2012, Cappellari et al.\ 2012).
vDC10 showed
that some spectral features, such as the \ion{Na}{1}
$\lambda\lambda8183,8195$ doublet (called NaI0.82 by CvD12), depend
strongly on surface gravity at fixed effective temperature, betraying
the presence of faint M dwarfs in integrated light spectra.  If
correct, the low-mass end of the IMF can be inferred directly from
red/near-IR spectra of old populations.  
Hence, the strength of the \ion{Na}{1} doublet versus another sodium
feature, such as the NaD doublet (called Na0.59 by CvD12), should
provide a powerful means for separating the IMF from other
effects. Specifically for the purpose 
of determining the low-mass IMF
down to $\sim 0.1M_{\odot}$ for metal-rich stellar populations with
ages of 3--13.5 Gyr, Conroy \& van Dokkum (2012; hereafter CvD12)
presented new population synthesis models. 
The NaD feature responds more strongly to Na-enhancement
than IMF in the CvD12 models, while the \ion{Na}{1} doublet is
strong in stars with mass $<0.3\,M_{\odot}$ and weak or absent in
all other types of stars. Unfortunately, NaI0.82 is also sensitive to age
and metallicity, and NaD is influenced by any interstellar
medium. It is therefore necessary to test these models over a
range of age and metallicity indicators, as well as against other 
lines caused by low-mass stars. 

In this letter, we focus on the NaI feature as indicator of low-mass
stars. We use  NaD as indicator of a change in sodium abundance and
H$\beta$ and [MgFe] as indicators of age and metallicity, respectively. 
This allows us to assess 
model degeneracies and deficiencies. We propose the use of the TiO feature 
at $\lambda$6230 as an indicator of the presence of low-mass stars.  
We find that both of these features (NaI and TiO) correlate with galaxy velocity dispersion, 
implying a steepening of the IMF slope in ETGs with $\sigma>\sigma_{*}$.
We assume $H_{\rm 0}=70 \,\mathrm{km \, s^{-1}\,Mpc^{-1}}$,
$\Omega_{\rm m}=0.3$ and $\Omega_{\Lambda}=0.7$ throughout this letter.

\begin{figure}  
\center
\includegraphics[height=12 cm]{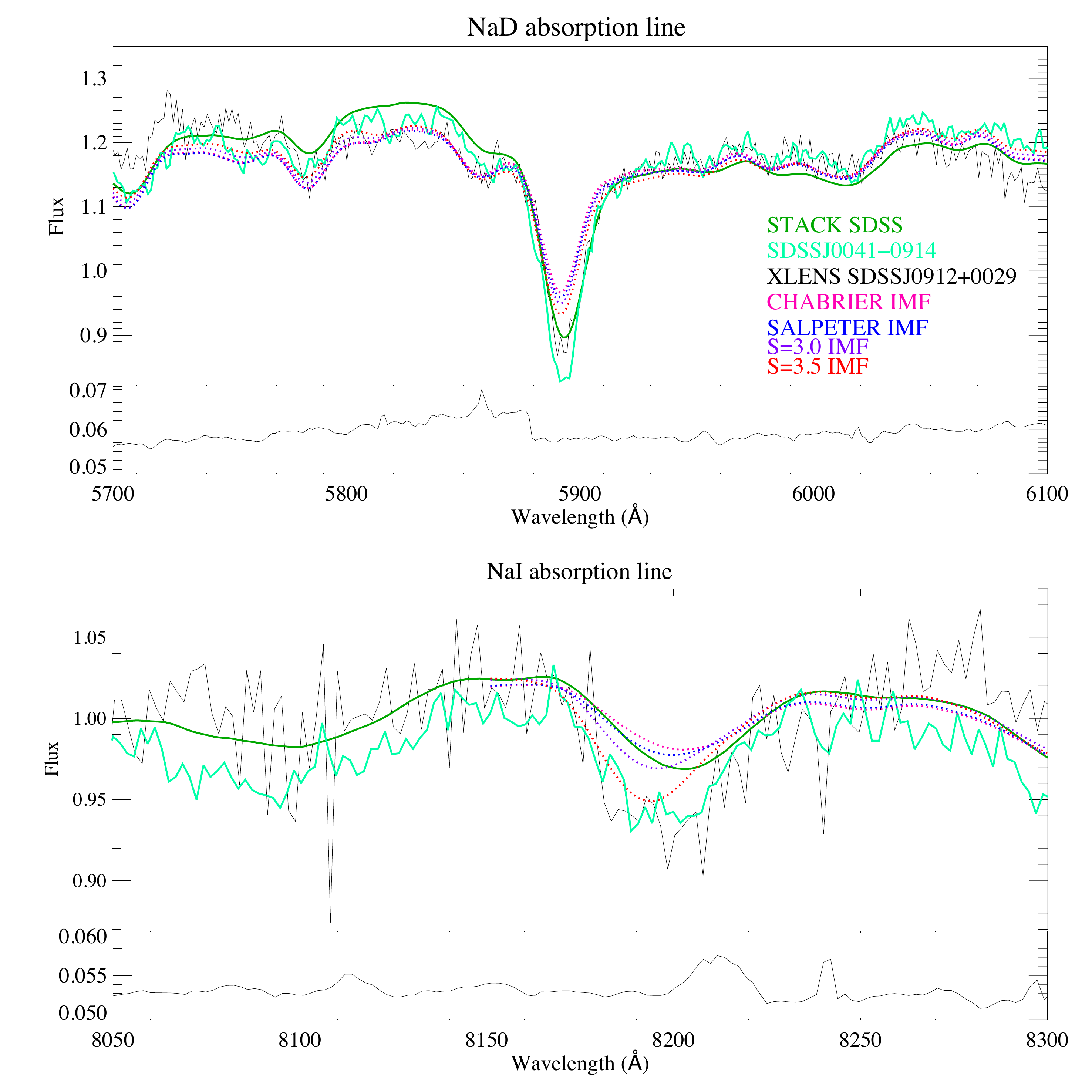}%{spectra_new_1c}%
\caption{Galaxy (continuous lines) and model (dashed lines) spectra in
  the regions of the NaD (top) and NaI (bottom) features.  The observed 
  NaD EWs do not match the models for the most massive ETGs ($\ge 300$\,\kms).  
  \ion{Na}{1} absorption is stronger in the XLENS system and SDSSJ0041-0914 and appears in
  both cases to require an IMF steeper than Salpeter, while the
  stacked SDSS spectrum shows a weaker \ion{Na}{1} feature that
  matches a model with a Salpeter IMF. The bottom panels show
  the noise spectrum of the XLENS system.}
\label{fig:spectra}
\end{figure}

\section{The data}

As part of the {\tt XLENS}\footnote{The X-Shooter Lens Survey,
Spiniello et al.\ (2011)} project, we obtained a UVB-VIS X-shooter
spectrum of the massive and luminous early-type SLACS (Sloan Lens ACS
Survey, Bolton et al. 2006) lens galaxy SDSS J0912+0029 at $z=0.1642$,
with high enough signal-to-noise to perform stellar population
analyses.  The lens galaxy shows a surprisingly deep NaI0.82 feature
(Fig.\ref{fig:spectra}), making it an extremely interesting target for
studying the low-mass end of the IMF in ETGs.  We measure the luminosity-weighted
velocity dispersion of the lens galaxy from the reduced
flux-calibrated 1D UVB--VIS spectrum using the Penalized Pixel Fitting
(pPXF) code of Cappellari \& Emsellem (2004). We obtain $\langle
\sigma_{*}\rangle(\la R_{\rm eff})=325\pm10\pm12\,\kms$, in
agreement with the previously published value ($\sigma \simeq 313 \pm
12\,\kms$; Bolton et al.\ 2006). We also used  the spectra of $\sim$250 galaxies 
with similar morphology and colors (all LRGs) from the Sloan Digital Sky
Survey DR8 (SDSS; Aihara et al.\ 2011), in five velocity-dispersion bins spread over 
200--335\,\kms ($\sim$50 galaxies per bin).  We examine one system, SDSSJ0041-0914, separately 
because it has a NaI0.82 feature comparably deep to the XLENS system.

\section{Stellar Population Synthesis Modeling}

We use the synthetic spectra of CvD12 to analyze the stellar
populations of these galaxies.
The models make use of two separate empirical libraries, the MILES
library covering 3500--7400\,\AA\ (S\'{a}nchez-Bl\'{a}zquez et
al.\ 2006) and the IRTF library of cool stars covering 8100--24000\,\AA\ 
(Cushing et al.\ 2005; Rayner et al.\ 2009).  They also
incorporate synthetic spectra with the purpose of investigating
changes in the overall metallicity or changes in the abundances of
individual elements and to cover the gap in wavelength between the two
empirical libraries.  We refer to CvD12 for details.
The abundance variations of single elements are
implemented at fixed [Fe/H], which
implies that the total metallicity $Z$ varies from model to model.
 We measure
line-strength indices in the range 4000--8400\,\AA, including the
standard Lick indices H$\beta$, Mg$b$, Fe5270, Fe5335, NaD and
a TiO index (TiO2) using the definitions of Trager et al. (1998),  and the commonly-used [MgFe]
combination\footnote{$\mathrm{[MgFe]} = \sqrt{(\mathrm{Fe5270} +
    \mathrm{Fe5335})/2 \times \mathrm{Mg}b}$, Gonz\'{a}lez (1993)}.
    
We define a modified index
around the \ion{Na}{1} doublet $8183,8195$\,\AA,
which seems to be strongly dependent on 
the low-mass end of the IMF (Table~1).  This index is slightly
different from that used by vDC10 and CvD12, having a 
wider central index bandpass and slightly wider
  pseudo-continua.  Our definition is more stable against velocity
dispersion variations and more suitable for massive
ETGs.
We convolve all the galaxy and model spectra to an effective
velocity dispersion of $\sigma=335\,\kms$\ (the upper limit in our
sample), to correct for kinematic broadening, before measuring indices.  Indices
in both the observed and synthetic spectra are measured with the same
definitions and method (SPINDEX2; Trager et al.\ 2008). We do not
place our indices on the zero-point system of the Lick indices and 
quote them as equivalent widths (EWs) in units of \AA, except for TiO2, which is given in magnitudes.
\\

\begin{deluxetable}{lcc}
\tablewidth{0pt}
\tablecaption{Definition of the index around the \ion{Na}{1} doublet $8183,8195$\AA \label{tab:sodium}}
\tablecolumns{3}
\tablehead{\colhead{   Index   }  & \colhead{   Central band   }& \colhead{   Pseudocontinua}\\
& \colhead{(\AA)} & \colhead{(\AA)}}
\startdata
NaI		& 		8168.500  -- 8234.125 	&  8150.000  -- 8168.400  \\
				&&  8235.250  -- 8250.000
\enddata \\  
\end{deluxetable}

\begin{figure}  
\includegraphics[height=11.2cm]{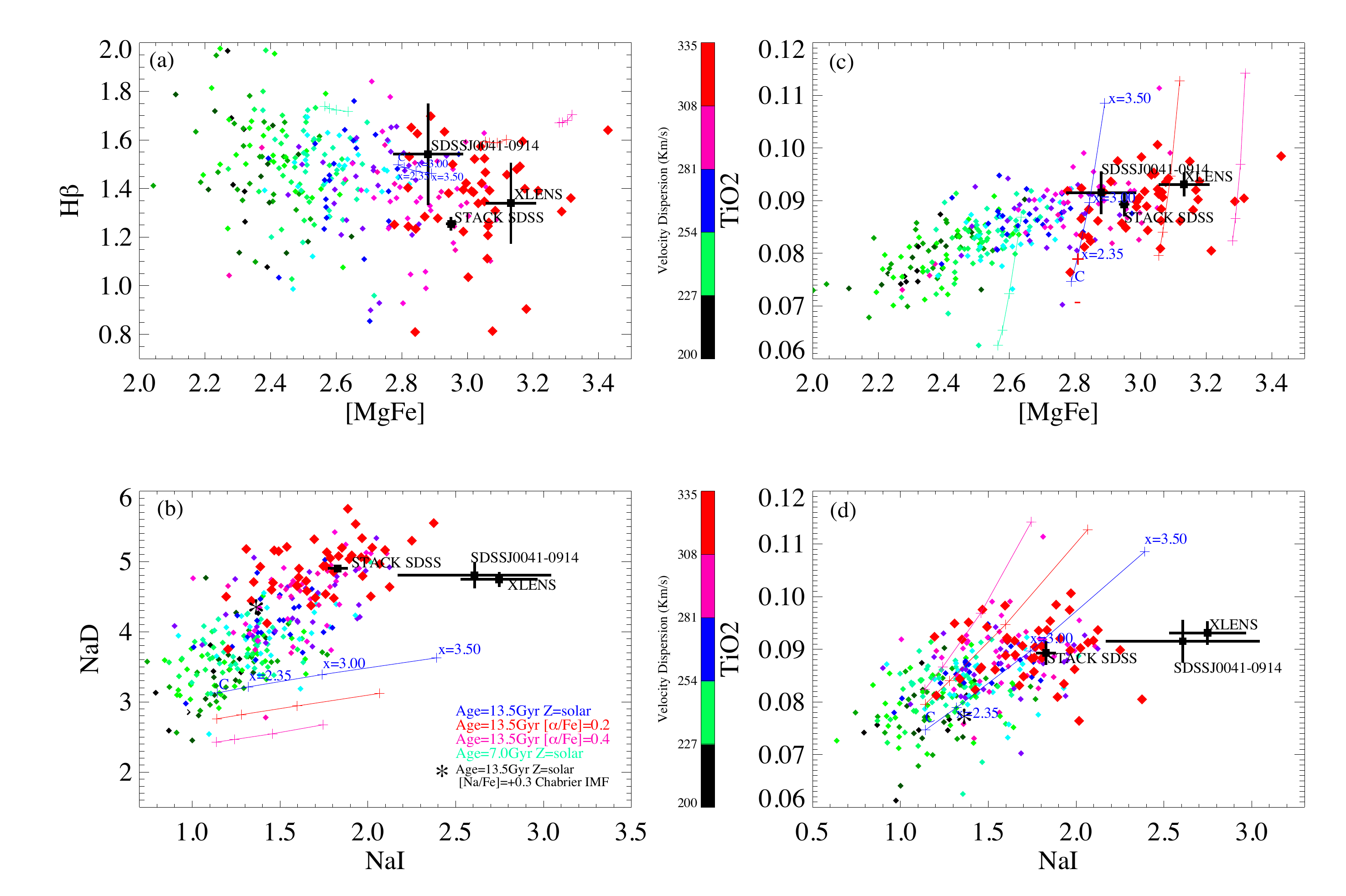}%{indices_new.pdf}%{f2.ps}
\caption{Index-index plots of the main absorption features. Lines and
  crosses are different SSP models from CvD12 with increasing IMF
  (Chabrier, Salpeter with a slope of $x=2.35$, a bottom-heavy IMF
  with slope of $x=3.0$ and an extremely dwarf-rich IMF with a slope
  of $x=3.5$). Points colored according to their velocity dispersions are individual SDSS galaxies, 
  with index errors similar to SDSS J0041-0914.  In the
  plots showing sodium, the XLENS system SDSSJ0912+0029 requires a
  very steep IMF, violating lensing contraints on its total mass (see
  the text for further details).  \textit{Panel (a):} H$\beta$ as a
  function of [MgFe]. The most massive ETGs ($>300$\,\kms) best match an old stellar 
  population (13.5 Gyr) with super-solar total metallicity. Lower-mass systems are younger. 
  \textit{Panel (b):} NaD as a function of NaI. Only low-mass ($<250$\,\kms)   systems match the models. More massive ETGs require a higher
  [Na/Fe] and the XLENS system and SDSSJ0041-0914 also require a
  very steep IMF slope.  \textit{Panel (c):} TiO2 as a function
  of [MgFe]. The most massive ETGs require an IMF slope slightly steeper
  than Salpeter.  A Chabrier-type IMF systematically underestimates
  the SDSS TiO2 EWs.  \textit{Panel (d):} TiO2 as a function
  of NaI. The ETGs match with the models using a Salpeter or slightly
  steeper IMF, but the XLENS system and SDSSJ0041-0914 still 
  do not match the SSP models well.}
\label{fig:indices}
\end{figure}

\section{Results and discussion}

H$\beta$ is primarily an age indicator,
while a combination of Mg$b$, Fe5270, and Fe5335 yields information on
the mean metallicity [Z/H] of the population (Worthey 1994) while
minimizing the effects of abundance ratio variations (e.g.,
Gonz{\'a}lez 1993; Trager et al.\ 2000a).  These indices (Panel (a),
Fig.~\ref{fig:indices}) show a good
agreement between the models and the galaxies EWs for old stellar
populations, with an age of $13.5 \pm 3$ Gyr for $\sigma \ge 300$\,\kms\ (black points)
and younger ages for lower mass ETGs.  The statistical error
is deduced directly from variations in H$\beta$ \footnote{For stellar
  populations with ages $>10$ Gyr, an uncertainty of 0.1 \AA\ in
  H$\beta$ corresponds to 1 Gyr uncertainty in the age (cf.\ Worthey
  1994).}.  The most massive ETGs have values of $[\alpha/\mathrm{Fe}]$ between
solar and super-solar ($\sim$0.2), in good agreement with the
prediction that massive galaxies have significantly super-solar
abundance ratios because of rapid, high-efficiency star formation
(Trager et al.\ 2000b; Thomas et al.\ 2005, Spolaor et
al.\ 2009, 2010).  Given the uncertainties in the line-strengths of
the two individual galaxies (SDSSJ0912+0029 and SDSSJ0041-0914),
we are unable to determine their ages and metallicities precisely, but
their line strengths are similar to the mean of the highest-mass SDSS sample,
with a deviation from the average EW smaller than $1\sigma$ in both
age and metallicity.
The NaI and NaD indices can in principle be used to
constrain the IMF slope (CvD12), and this relation is shown in Figure 2b.  Although the data match
the models for low-dispersion systems ($\la$250\kms), the models with solar [Na/Fe] abundance do not match
the NaD strengths and only models with $\mathrm{[Na/Fe]}=+0.3 \,-\,+0.4$ dex
match the NaD indices for higher-mass ETGs.  
We suggest two possible explanations for this behavior:  

(i) NaD is highly contaminated by the
interstellar medium (ISM) for higher mass ETGs; for example, dust lanes provide additional
absorption in this resonance line (Sparks et al.\ 1997).  
Interstellar absorption within a galaxy may alter the stellar absorption 
profile and therefore the calculated EW, leading to an incorrect inference of the underlying stellar population.

(ii) Very massive ETGs have higher [Na/Fe] abundances ($>0.3$
dex) \emph{and} slightly bottom-heavy IMFs which correlate with their stellar velocity dispersions.
Therefore, if we explain the strengths of these features in giant ETGs using abundance ratios, 
we require an average iron abundance in excess of solar  ([Fe/H] $\sim 0.2$), a IMF with $x=3.0$  
and a high sodium abundance ([Na/H] $> 0.3$).
In the $\alpha-$enhanced bulge of the Galaxy, Fulbright et al. (2006)
find an averaged [Na/Fe]=0.2 dex, and that [Na/Fe] $\le 0.3$ dex
in all stars.
However, Lecureur et al. (2007) find that [Na/Fe] ratios increase sharply with metallicity. 
They obtain values of [Na/Fe] $\sim 0.5$ for [Fe/H] = 0 and even higher for  [Fe/H] $>0$, 
but with a scatter of 0.29 dex resulting in a range of [Na/Fe] from $-0.1$ to almost $1.0$.
It is therefore possible for massive ellipticals to have high [Na/Fe]. 
In both cases, the models seem
consistent with a Salpeter IMF at the low-dispersion end and a slightly bottom-heavy IMF for the high-dispersion end, 
if these effects are accounted for, but the models predict a steeper IMF slope of $x\sim3.0-3.5$ for both the
XLENS galaxy SDSSJ0912+0029 and SDSSJ0041-0914. 

We note that a TiO feature at 8199\AA\ could partly contaminate NaI, although this feature should
not vary strongly (CvD12).  To test possible contamination, we use a model with [Ti/Fe]=$\pm 0.3$ and  
calculate the NaI EW for a Chabrier IMF. We find that Ti enhancement only affect the NaI index by 1\%. 
%Ti/Fe = 0.3  Chabrier IMF, EW(NaI) = 1.1221e+00
%Ti/Fe = 0.0  Chabrier IMF, EW(NaI) = 1.1418e+00
%delta = 0.0197 -> 1%!!! 

Overall we conclude that the NaD EWs and its trend with stellar mass remain unexplained for 
systems with $\sigma \ga 250$\,\kms.
We find that SSP models predict that TiO features 
also depend strongly on the slope of the low-mass end of the IMF, such as TiO2, shown in Figures 2c and 2d.
This indicator gives more support to the conclusion that the sodium strengths of the
XLENS ETG, SDSSJ0912+0029, still remain somewhat difficult to explain by current
stellar population models, although most SDSS systems can be matched in NaI
for most ETGs (if not in NaD).  Together the TiO2 and NaI indices both
imply a bottom-heavy IMF, steepening from Salpeter to possibly $x \approx 3$
for the most massive SDSS ETGs.  As in Treu et al.\ (2010), a
bottom-light IMF such as Chabrier IMF is inappropriate for the most massive ETGs.

\begin{deluxetable}{lcccc}
\rotate
\tablewidth{0pt}
\tablecolumns{3}
\tablecaption{Variation with IMF of M/L and stellar mass fraction within the Einstein radius\label{tab:m2l}}
\tablehead{\colhead{IMF slope}  & \colhead{$(M/L)^{*}_{DSEP,B}$} & \colhead{$(M/L)^{*}_{DSEP,V}$} &\colhead{$f^{*}_{B}$} &\colhead{$f^{*}_{V}$}\\
\colhead{($dN/dm=M^{\gamma}$)} & \colhead{$([\alpha/Fe]=0.0)$} & \colhead{$([\alpha/Fe]=0.0)$} & \colhead{$(L_{\mathrm{Ein}}/M_{\mathrm{Ein}}) \times (M/L)^{*}_{B}$} & \colhead{$(L_{\mathrm{Ein}}/M_{\mathrm{Ein}}) \times (M/L)^{*}_{V}$}  }
\startdata
$2.35$  & $  10.2 \pm 3$& $  7.2 \pm 2$  &  $0.75 \pm 0.2$&  $0.59 \pm 0.18$\\
$3.00$  & $  22 \pm 6$   & $16 \pm 5$  & $1.6 \pm 0.5$&  $  1.4 \pm 0.4$\\
$3.50$  & $  43 \pm 13$ & $29 \pm 9$  &  $2.4 \pm 0.8$& $  2.4 \pm 0.7$
\enddata
\tablecomments{The constraints on the mass and luminosity within the Einstein radius are taken from Auger et al.\ (2009)
  and Barnab\`{e} et al.\ (2009). All quantities are calculated in the rest-frame $V$- and $B$-band.}
\end{deluxetable}

\subsection{Limits on the IMF from Strong Lensing}

A strong case against an extreme bottom-heavy IMF can be made using
the system with the strongest NaI EW (Fig.~\ref{fig:indices}), the
XLENS galaxy SDSSJ0912+0029. This system provides a hard upper limit
on the stellar mass inside its Einstein radius, no matter the IMF model.
If we assume that the SSP models are correct and that this galaxy has a
high [Na/Fe] abundance, we infer an IMF with a
power-law slope $x=3$--$3.5$ (where the IMF follows
$dn/dm=m^{-x}$, and the Salpeter slope is $x=2.35$).  To assess
whether these steep IMF slopes are consistent with the
upper limit on the total mass, we calculate the total luminosity and the SSP 
stellar M/L ratio in stars for each assumed IMF to infer the stellar
mass fraction inside the Einstein radius
($R_{\rm Ein}=4.55 \pm 0.23$\,kpc; Koopmans et al.\ 2006).
Changes in the IMF of stars with $M \leq 0.3\,M_{\odot}$ changes the
total luminosity of the lens galaxy by at most $\sim 10\%.$
Conversely,  
stars with masses of $0.1$--$0.3\,{M}_{\odot}$ contribute $\ga 60\%$
of the stellar mass for bottom-heavy IMFs with slopes steeper
than Salpeter (see, e.g., Fig.~2 of CvD12).  To determine the stellar
M/L ratio, we use the isochrones at solar [Fe/H] and
[$\alpha$/H] for a 13.5 Gyr population from the Dartmouth Stellar
Evolution Program (DSEP), a state-of-the-art stellar evolution code 
(Chaboyer, Green, \& Liebert 1999; Chaboyer et al.\ 2001).  We compare
three different IMFs: Salpeter ($x=2.35$), a bottom-heavy IMF
($x=3.0$) and a very bottom-heavy IMF ($x=3.5$). CvD12 use the same isochrones in their SSP
for the bulk of the main sequence and red giant branch, except at $M <
0.2\,M_{\odot}$, where they use the Baraffe et al.\ (1998)
  isochrones.
For each IMF we compute the quantity
\begin{equation*}
 f^{*}_{\mathrm{Ein}}=M^{*}/M_{\mathrm{Ein}} =
 (L_{\mathrm{Ein}}/M_{\mathrm{Ein}}) \times (M_{*}/L)_{\mathrm{DSEP}},
\end{equation*}
where $M_{\mathrm{Ein}}$ is a robust measurement of the total mass
enclosed within the physical Einstein radius [$M_{\mathrm{Ein}} =
(39.6 \pm 0.8) \times 10^{10}\,M_{\odot} \,$], $L_{\mathrm{Ein}}$ is
the luminosity enclosed within the Einstein radius, evaluated using
B-spline luminosity models, as a fraction of de Vaucouleurs total
model luminosity [$L_{\mathrm{Ein}} = (4.49 \pm 0.2) \times
10^{10}\,L_{\odot} \,$, from Bolton et al.\ 2008], and
$(M_{*}/L)_{\mathrm{DSEP}}$ is the mass-to-light ratio from the DSEP
isochrone using the appropriate IMF.  
The stellar M/L ratio includes the contribution from stellar remnants and gas 
ejected from stars at the end of their life-cycles.
We list the results of this calculation in Table~\ref{tab:m2l}.  For a
Salpeter IMF, the stellar mass fraction of SDSSJ0912+0029 in the
restframe $V$-band is $f^{*}_{\mathrm{Ein,Salp}}=0.59 \pm 0.15$, in
agreement with previous results ($0.60\pm0.09$, Auger et
al.\ 2009).  The mass-to-light ratio calculated from the DSEP
isochrone for a Salpeter IMF is $M/L_V =7.2 \pm 2\,(M/L)_{\odot}$ in
the $V$ band and $M/L_B =10.2 \pm 3\,(M/L)_{\odot}$ in the $B$ band.
The latter value is consistent with the upper limit of
$M/L_B\leq9.08\,(M/L)_{\odot}$ derived from dynamical models of
Barnab\`{e} at al.\ (2009) under the maximum bulge hypothesis.
An IMF slope of
$x=3.5$ yields $M/L_{V} =29 \pm 9\,(M/L)_{\odot}$, and $M/L_{B} =43
\pm 13\,(M/L)_{\odot}$ corresponding to
$f^{*}_{\mathrm{Ein,3.5}}=2.4\pm0.8$, inconsistent with the total
lensing mass within the Einstein radius at the $>95$\% confidence
level. An IMF of $x=3.0$ in $B$-band is also excluded at the $>90$\%
level, as this corresponds to a fraction
$f^{*}_{\mathrm{Ein,3.0},B}=1.6\pm0.5$.
For both of the bottom heavy IMFs in $B$-band and for the $x=3.5$ IMF
in $V$-band, we obtain a stellar mass fraction within the
Einstein radius in excess of unity, thereby violating the
lensing constraint on the total mass of the system at the $>90$\%
CL. The $x=3$ model is only marginally consistent in $V$-band, but
$f^{*}_{\mathrm{Ein,3.0},V}=1.4\pm0.4$ implies that 
there is no dark matter within the Einstein radius.

\subsection{Systematic uncertainties}

The uncertainty on the value of $f^{*}_{\mathrm{Ein}}$ has a number of contributions. 
The uncertainties in the mass and luminosity determinations from lensing
are much smaller than differences in the values of $M/L$ arising from
the use of different stellar population evolution models.
The emerging picture is
that, for a fixed IMF, it is difficult to constrain $M/L$ estimates to
much higher accuracy than 0.1 dex (Gallazzi et al.\ 2008; Marchesini et al.\ 2009; Longhetti \& Saracco
2009, Conroy et al.\ 2009, 2010).
We examine mass-to-light ratios predicted for different IMF from different
stellar population models in the rest-frame $V$- and $B$-bands and 
compare predictions from Worthey (1994), Bruzual \& Charlot (2003),
Maraston (2005), and Vazdekis et al.\ (2010) for single stellar
populations with ages 11.2--14.1\,Gyr, solar ($Z=0.02$) or
super-solar metallicity ($Z=0.05$).  For each SSP and each IMF, we
calculate an average value and a standard deviation that we
associate with the inferred values of $M/L$.  
Changing the [Fe/H] abundance from $0$ to $0.22$ yields a $\sim 9\%$
uncertainty on $M/L$, while changing the age of the stellar population 
changes $M/L$ by $\sim20\%$ at fixed IMF.  The latter is the
dominant contribution to the final uncertainties. We propagated these errors 
into the stellar mass fraction. The errors on the stellar mass fractions in Table~\ref{tab:m2l}
include both the random error contribution and the systematic
uncertainties due to the use of different set of isochrones, bands,
and stellar population age and metallicity uncertainties.

\section{Conclusions}

In this letter we have studied the \ion{Na}{1} $\lambda$8190 
and TiO $\lambda$6230 features -- both indicators of low-mass ($<
0.3\,M_{\odot}$) stars in massive ETGs -- as a function of each other, 
of age and metallicity indicators (Mg$b$, Fe, H$\beta$), of NaD, and of stellar velocity dispersion. 
We find the following: 
(1) The observed NaI-NaD trend depends
strongly on stellar velocity dispersion of ETGs and only match
current state-of-the-art SSP models for ETGs with $\sigma \la 250$\,\kms.
The most extreme NaI index strength in our sample is found in a 
gravitational lens system, which should have an IMF slope $x \ga 3$ based on 
the best current SSP models.  The total enclosed
mass of this system, however, excludes slopes steeper than $x=3.0$ at
the $>90\%$ CL or slopes steeper than $x=3.5$ at the $>95\%$ CL.
We conclude that the NaD feature is still affected by 
as-of-yet not understood processes in the more massive ETGs
($\sigma>250$\,\kms).  
A full spectral comparison, 
in combination with lensing and dynamical constraints, is planned to further strengthen these 
results and assess whether NaI and NaD (in some instances) are contaminated. 
(2) We find that the TiO feature at $\lambda\sim 6230$\AA\ (TiO2) 
is a particularly promising feature to decouple the IMF from age,
metallicity, and abundance pattern of the stellar population, especially when combined with
metallicity-dependent indices.  We find that this feature correlates well with 
NaI, if the two most extreme cases as discussed in the text, are excluded. 
This correlation can be a crucial piece of evidence against interstellar contamination
of the \ion{Na}{1} $\lambda$8190  sodium absorption lines, although this does not solve the problem of NaD absorption.
If strong NaI features are indeed not due to ISM contamination, very massive ETGs have higher [Na/Fe] abundances ($>0.3$
dex) \emph{and} slightly bottom-heavy IMFs, correlated with their stellar velocity dispersions.
We also find a clear trend of an increasing IMF slope between $\sigma = 200$ to 335\,\kms\ from 
Salpeter ($x=2.35$) to $x\approx 3.0$, in agreement with the XLENS system, which excludes 
steeper IMFs at the high-mass end.  

Our results are the first SSP-based indications of a steepening of the low-mass end of the IMF 
with increasing galaxy mass {\sl within} the class of LRG/ETGs. Our results (i) support a similar 
trend first found by \citet[][]{2010ApJ...709.1195T}, (ii) extend the evidence based 
on SSP models  that the IMF steepens from spiral to early-type 
galaxies (vDC10), (iii) suggest that NaI and NaD (in some instances) could be contaminated by interstellar absorption,
and (iv) support a similar trend found by 
\citet[][]{2012arXiv1202.3308C}  based on stellar kinematics.  
The upper limit of  $x\la 3$, based on one of the most massive  ETGs in our sample,
a gravitational lens, also supports our previous similar finding \citep[][]{2011MNRAS.417.3000S}.

\section*{Acknowledgements}
The authors thank the referee for providing constructive comments.
Data were reduced using EsoRex and the XSH pipeline developed by the
ESO Data Flow System Group. C.S. acknowledges support from an Ubbo
Emmius Fellowship. L.V.E.K. is supported in part by an NWO-VIDI
program subsidy (project number 639.042.505).  The authors thank
C. Conroy and P. van Dokkum for kindly providing their stellar
population models before publication and for providing
very useful feedback on a draft manuscript that helped to improve
it. The authors thank T. Treu and M. den
Brok for useful comments on the manuscript.

%\clearpage

\end{document}